\documentclass[12pt]{article}

\usepackage{amsmath}
\usepackage{amssymb}

\usepackage{graphicx}

\usepackage{cite}

\usepackage[labelfont=bf,labelsep=period,justification=raggedright]{caption}
\usepackage{caption}

\makeatletter
\renewcommand{\@biblabel}[1]{\quad#1.}
\makeatother

\date{}

\begin{document}

\begin{flushleft}
{\Large
\textbf{Google matrix analysis of DNA sequences}
}
\\ \bigskip
Vivek\ Kandiah$^{1}$, 
Dima L.\ Shepelyansky$^{1,*}$
\\ \medskip
{1} {\it Laboratoire de Physique Th\'eorique du CNRS, IRSAMC, 
Universit\'e de Toulouse, UPS, F-31062 Toulouse, France}
\\
\medskip
$\ast$ Webpage: www.quantware.ups-tlse.fr/dima
\end{flushleft}

\section*{Abstract}

For DNA sequences of various species we construct
the Google matrix $G$ of Markov transitions 
between nearby words composed of several letters.
The statistical distribution of matrix elements
of this matrix is shown to be described by a power law
with the exponent being close to those of outgoing links in
such scale-free networks as the World Wide Web (WWW).
At the same time the sum of ingoing matrix elements
is characterized by the exponent being significantly
larger than those typical for WWW networks.
This results in a slow algebraic decay of the PageRank
probability determined by the distribution of 
ingoing elements. The spectrum of $G$ is characterized by a 
large gap leading to a rapid relaxation process on the
DNA sequence networks. We introduce the PageRank proximity 
correlator between different species 
which determines their statistical similarity
from the view point of Markov chains.
The properties of other eigenstates of the Google
matrix are also discussed. Our results establish 
scale-free features of DNA sequence networks
showing their similarities and distinctions 
with the WWW and linguistic networks.\\

\noindent
Dated: January 8, 2013

\newpage$\phantom{.}$

\section*{Introduction}

The theory of Markov chains \cite{markov} 
finds impressive modern applications to information retrieval
and ranking of directed networks including
the World Wide Web (WWW) where the number of nodes
is now  counted by tens of billions.
The PageRank algorithm (PRA) \cite{brin}
uses the concept of the Google matrix $G$
and allows to rank all WWW nodes in an efficient
way. This algorithm is a fundamental element
of the Google search engine used by a majority of Internet users.
A detailed description of this method 
and basic properties of the Google matrix
can be found e.g. in \cite{meyerbook,ukuniv}.

The Google matrix belongs to the class of Perron-Frobenius operators
naturally appearing in dynamical systems (see e.g. \cite{mbrin}).
Using the Ulam method \cite{ulam} a discrete approximant
of Perron-Frobenius operator can be constructed for simple
dynamical maps following only one trajectory in a chaotic
component \cite{ulamfrahm} or  using many independent
trajectories counting their probability transitions
between phase space cells \cite{froyland,sz},\cite{acat}. 
The studies of Google matrix 
of such directed Ulam networks provides an interesting
and detailed analysis of dynamical properties of 
maps with a complex chaotic dynamics \cite{ulamfrahm,froyland},\cite{sz,acat}.

In this work we use the Google matrix approach to
study the statistical properties of DNA sequences
of  the species:  Homo sapiens (HS, human), 
Canis familiaris (CF, dog), Loxodonta africana (LA, elephant),
Bos Taurus (bull, BT),  Danio rerio (DR, zebrafish),
taken from the publicly available database
\cite{genbank}. The analysis of Poincar\'e recurrences
in these DNA sequences \cite{dnafrahm} shows their
similarities with the statistical properties of
recurrences for dynamical trajectories in 
the Chirikov standard map and other symplectic maps \cite{ulamfrahm}. 
Indeed, a DNA sequence can be 
viewed as a long symbolic trajectory
and hence, the Google matrix, constructed from it,
highlights the statistical features of DNA
from a new viewpoint.  

An important step in the statistical analysis of DNA sequences
was done in \cite{buldyrev1995} applying methods of
statistical linguistics and determining the frequency of 
various words composed of up to 7 letters.
A first order Markovian models have been also
proposed and briefly discussed in this work.
Here we show that the Google matrix analysis
provides a natural extension of this approach.
Thus the PageRank eigenvector gives
the frequency appearance of words of given length.
The  spectrum and eigenstates of $G$ characterize the relaxation
processes of different modes in the Markov process
generated by a symbolic DNA sequence.
We show that the comparison of word ranks of different species
allows to identify proximity between species.

At present the investigations of statistical properties of DNA
sequences are actively developed by various bioinformatic  groups
(see e.g. \cite{robinbook,robinplos},\cite{wang},\cite{reinert,burden}).
The development of various methods of statistical analysis 
of DNA sequences become 
now of great importance due to a rapid growth of collected genomic data.
We hope that the Google matrix approach, which already demonstrated its 
efficiency for enormously large networks \cite{brin,meyerbook}, will find 
useful applications for analysis of genomic data sets. 

\section*{Results}

\subsubsection*{Construction of Google matrix from DNA sequence}
\label{ssec1}\ 
From \cite{genbank} we collected DNA sequences
of HS represented as a single string of length
$L \approx 1.5 \cdot 10^{10}$ base pairs (bp)
corresponding to 5 individuals.
Similar data are obtained for  BT  ($2.9 \cdot 10^9$ bp),
CF ($2.5 \cdot 10^9$ bp), LA  ($3.1 \cdot 10^9$ bp), DR ($1.4 \cdot 10^9$ bp).
For HS, CF, LA, DR the statistical properties
of Poincar\'e recurrences in these sequences 
are analyzed in \cite{dnafrahm}.
All strings are composed of 4 letters $A,G,G,T$
and undetermined letter ${\it N_l}$. 
The strings can be found at 
the web page \cite{dnagoogle}.

For a given sequence we fix the words $W_k$ of $m$ letters length
corresponding to the number of states $N=4^m$.
We consider that there is a transition 
from a state $i$ to state $j$ inside this basis $N$
when  we move along the string from left to right
going from a word $W_k$ to a next word $W_{k+1}$.
This transition adds one unit in the transition
matrix element $T_{ij} \rightarrow T_{ij}+1$.
The words with letter ${\it N_l}$
are omitted, the transitions are counted only between nearby words
not separated by words with $N_l$.
There are approximately $N_t \approx L/m$
such transitions for the whole length $L$
since the fraction of undetermined letters ${\it N_l}$ is small.
Thus we have $N_t=\sum_{i,j=1}^{N} T_{ij}$.
The Markov matrix of transitions $S_{ij}$ is
obtained by normalizing matrix elements in such a way that their sum
in each column is equal to unity: $S_{ij}=T_{ij}/\sum_i T_{ij}$.
If there are columns with all zero elements (dangling nodes)
then zeros of such columns are replaced by $1/N$.
Such a procedure corresponds to one used 
for the construction of Google matrix of the WWW \cite{brin,meyerbook}.
Then the Google matrix of DNA sequence is written as
\begin{equation}
\label{eq1}
G_{ij}=\alpha S_{ij}+(1-\alpha)/N ,
\end{equation}
where $\alpha$ is the damping factor
for which the Google search uses usually the value
$\alpha \approx 0.85$ \cite{meyerbook}.
The matrix $G$ belongs to the class of Perron-Frobenius operators.
It has the largest eigenvalue $\lambda=\lambda_1=1$
with all other eigenvalues $|\lambda_i| \leq \alpha$.
For WWW usually there are isolated subspaces 
so that at $\alpha=1$ there are many degenerate $\lambda=1$
eigenvalues \cite{ukuniv} 
so that the damping factor allows to eliminate this degeneracy
creating a gap between $\lambda=1$ and all other eigenvalues.
For our DNA Google matrices we find that there is already 
a significant spectral gap naturally present.
In this case the PageRank vector is not sensitive
to the damping factor being in the range
$0.5 \leq \alpha \leq 1$ 
(other eigenvectors are independent of $\alpha$ 
\cite{meyerbook,ukuniv},\cite{sz}).
Due to that in the following we present all results
at the value $\alpha=1$. 

The spectrum $\lambda_i$ and right eigenstates $\psi_i(j)$ are determined
by the equation
\begin{equation}
\label{eq1}
\sum_{j'} G_{jj'} \psi_i(j') =\lambda_i  \psi_i(j)  .
\end{equation}
The PageRank eigenvector $P(j)$ at $\lambda=1$ has positive or zero elements
which can be interpreted as a probability to find a random surfer
on a given site $j$ with the total probability normalized to unity
$\sum_j P(j)=1$. Thus, all sites can be ordered in a decreasing order
of probability $P(j)$ that gives us the PageRank order index
$K(j)$ with most frequent sites at low values of $K=1,2,...$.

\begin{figure*}[!ht] 
\begin{center} 
\includegraphics[width=0.9\columnwidth]{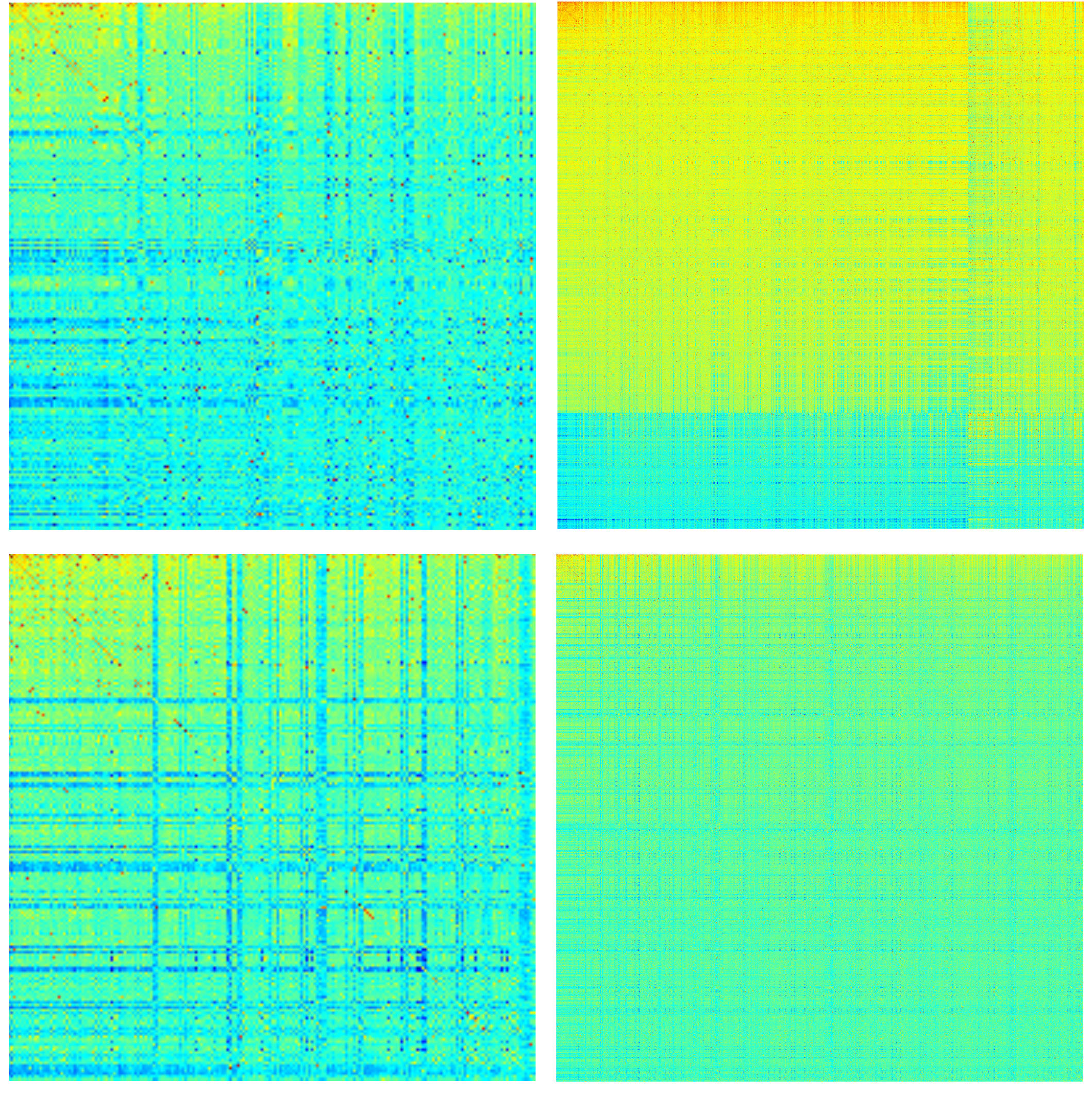}
\caption {\baselineskip 14pt
DNA Google matrix of Homo sapiens (HS) constructed for words of 5-letters (top) 
and 6-letters (bottom) length. Matrix elements $G_{KK'}$ are shown 
in the basis of PageRank index $K$ (and $K'$). Here, $x$ and $y$ axes show 
$K$ and $K'$ within the range $1 \leq K,K' \leq 200$ (left) and 
$1 \leq K,K' \leq 1000$ (right). The element $G_{11}$ at
 $K=K'=1$ is placed at top left corner.
Color marks the amplitude of 
matrix elements changing from blue for minimum zero value 
to red at maximum value.}
\label{fig1}\label{figure1} 
\end{center}
\end{figure*}

It is useful to consider the density of matrix elements 
$G_{KK'}$ in the PagePank indexes $K,K'$
similar to the presentation used in \cite{2dmotor,twitter}
for networks of Wikipedia, UK universities, Linux Kernel 
and Twitter. The image of the DNA Google matrix of HS
is shown in Fig.~\ref{fig1} for words of 5 and 6 letters.
We see that almost all matrix is full that 
is drastically different from 
the WWW and other networks considered in \cite{2dmotor}
where the matrix $G$ is very sparse.
Thus the DNA Google matrix is
more similar to the case of Twitter which is characterized
by a strong connectivity of top PageRank nodes \cite{twitter}.

\begin{figure*}[!ht] 
\begin{center} 
\includegraphics[width=0.9\columnwidth]{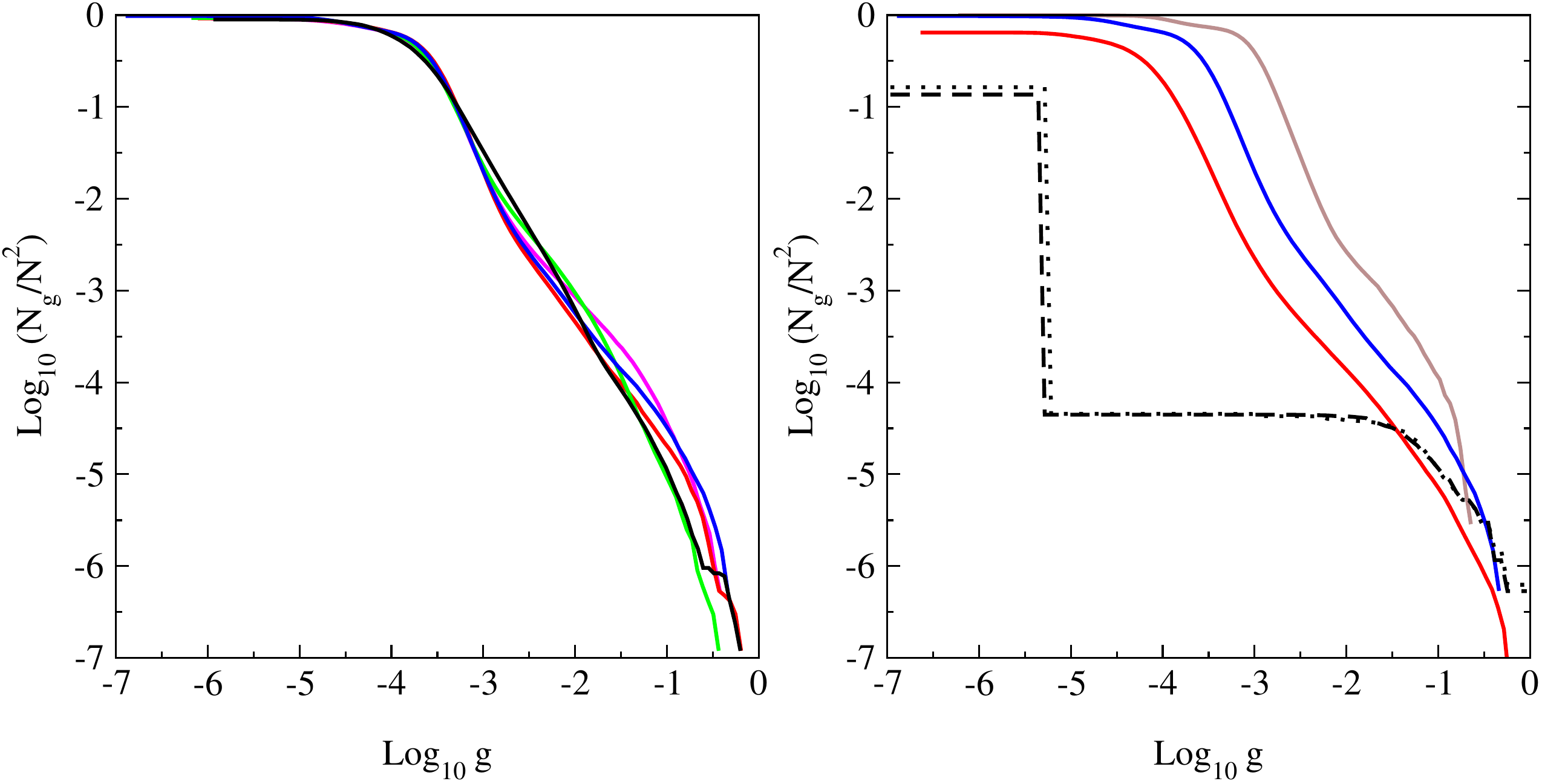}
\caption {\baselineskip 14pt
Integrated fraction $N_g/N^2$ of  Google matrix elements with $G_{ij} > g$ 
as a function of $g$.
\emph{Left panel :} Various species with 6-letters word length: 
bull BT (magenta), dog CF (red), elephant LA (green), 
Homo sapiens HS (blue) and zebrafish DR(black). 
\emph{Right panel :} Data for HS sequence with 
words of length $m= 5$ (brown), $6$ (blue), $7$ (red).
For comparison black dashed and dotted curves
show the same distribution for the WWW networks
of Universities of Cambridge and Oxford in 2006
respectively.}
\label{fig2}\label{figure2} 
\end{center}
\end{figure*}

It is interesting to analyze the statistical properties 
of matrix elements $G_{ij}$.
Their integrated distribution is shown in Fig.~\ref{fig2}. 
Here $N_g$ is the number of
matrix elements of the matrix $G$ with
values $G_{ij} > g$. The data show that the number 
of nonzero matrix elements $G_{ij}$ 
is  very close to $N^2$.
The main fraction of elements has values $G_{ij} \leq 1/N$
(some elements $G_{ij} < 1/N$ since 
for certain $j$ there are many  transitions
to some node $i'$ with $T_{i'j} \gg N$
and e.g. only one transition to other $i''$ with $T_{i''j}=1$).
At the same time there are also transition elements $G_{ij}$
with large values whose fraction decays
in an algebraic law $N_g \approx  A N/g^{\nu-1}$
with some constant $A$ and an exponent $\nu$. 
The fit of numerical data  in the range 
 $-5.5 < \log_{10}g < -0.5$ of algebraic decay
gives for $m=6$: $\nu=2.46 \pm 0.025$ (BT),
$2.57 \pm 0.025$ (CF),  $2.67 \pm 0.022$ (LA),
$2.48 \pm 0.024$ (HS), $2.22 \pm 0.04$ (DR).
For HS case we find
$\nu = 2.68 \pm 0.038$ at $m=5$ and
$\nu = 2.43 \pm 0.02$ at $m=7$ with the average  
$A \approx 0.003$ for $m=5,6,7$.
There are visible oscillations in the algebraic
decay of $N_g$ with $g$ but in global
we see that on average all species are 
well described by a universal decay law
with the exponent $\nu \approx 2.5$.
For comparison we also show the distribution
$N_g$ for the WWW networks of University of Cambridge
and Oxford in year 2006 (data from \cite{ukuniv,2dmotor}).
In these networks we have $N \approx 2 \cdot 10^5$
and on average 10 links per node. We see that in these
cases the distribution $N_g$ has a very short 
range in which the decay is at least 
approximately algebraic ($-5.5 < \log_{10}(N_g/N^2) < -6$).
In contrast to that for the DNA sequences
we have a large range of algebraic decay.

Since in each column we have the sum of
all elements equal to unity we can say that
the differential fraction
$d N_g/dg \propto 1/g^{\nu}$ gives the distribution of
outgoing matrix elements which is similar to the distribution
of outgoing links extensively studied for the WWW networks
\cite{meyerbook,donato},\cite{upfal,zzs}. 
Indeed, for the WWW networks all links in a column
are considered to have the same weight
so that these matrix elements are given by 
an inverse number of outgoing links \cite{meyerbook}. 
Usually the distribution
of outgoing links follows a power law
decay with an exponent $\tilde{\nu} \approx 2.7$
even if it is known that this exponent is much more
fluctuating compared to the case of ingoing links.
Thus we establish that the distribution of DNA matrix elements
is similar to the distribution of outgoing links
in the WWW networks with $\nu \approx \tilde{\nu}$.
We note that for the distribution of outgoing links
of Cambridge and Oxford networks
the fit of numerical data gives the
exponents
$\tilde{\nu} = 2.80 \pm 0.06$ (Cambridge) and
$2.51 \pm 0.04$ (Oxford).

\begin{figure*}[!ht] 
\begin{center} 
\includegraphics[width=0.9\columnwidth]{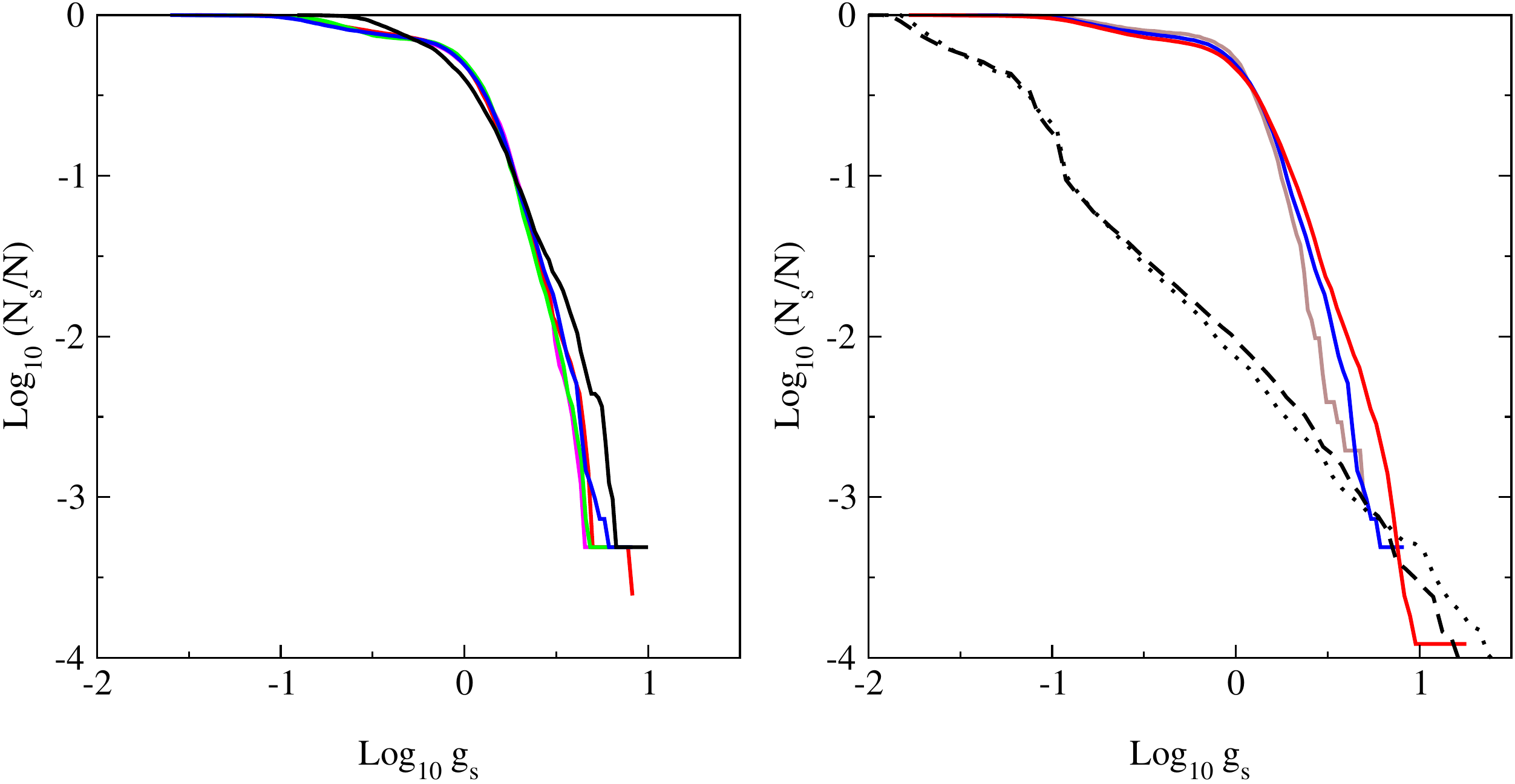}
\caption {\baselineskip 14pt
Integrated fraction $N_s/N$ of sum of ingoing matrix elements 
with $\sum_{j=1}^NG_{i,j} \ge g_s$. Left and right panels
show the same cases as in Fig.~\ref{fig2} in same colors.
The dashed and dotted curves are shifted in $x$-axis 
by one unit left to fit the figure scale.}
\label{fig3}\label{figure3} 
\end{center}
\end{figure*}

It is known that on average the probability of
PageRank vector is proportional to the number of
ingoing links \cite{meyerbook}. This relation
is established for scale-free networks
with an algebraic distribution of links
when the average number of links per node is 
about $10$ to $100$ that is usually
the case for WWW, Twitter and Wikipedia networks
\cite{ukuniv,2dmotor},\cite{twitter,wikispectr},\cite{donato,upfal},
\cite{zzs}.
Thus in such a case the matrix $G$ is very sparse.
For DNA we find an opposite situation where 
the Google matrix is almost full
and zero matrix elements are practically absent.
In such a case an analogue of number of ingoing
links is the sum of ingoing matrix elements
$g_s=\sum_{j=1}^N G_{ij}$. 
The integrated distribution of
ingoing matrix elements with 
the dependence of $N_s$ on $g_{s}$ is shown in Fig.~\ref{fig3}.
Here $N_s$ is defined as the number of nodes with the sum of ingoing 
matrix elements being larger than $g_{s}$.
A significant part of this dependence, corresponding to large
values of $g_s$ and determining
the PageRank probability decay,
is well described by a power law
$N_s \approx B N /g_{s}^{\mu-1}$.
The fit of data at $m=6$ gives
$\mu=5.59 \pm 0.15$ (BT), $4.90 \pm 0.08$ (CF),
$5.37 \pm0.07$ (LA), $5.11 \pm 0.12$ (HS),
$4.04 \pm 0.06$ (DR).
For HS case at $m=5,7$ we find respectively
$\mu=5.86 \pm 0.14$ and $4.48 \pm 0.08$.
For $HS$ and other species we have an average 
$B \approx 1$.

Usually for ingoing links distribution
of WWW and other networks 
one finds the exponent $\tilde{\mu} \approx 2.1$
\cite{donato,upfal},\cite{zzs}.
This value of $\tilde{\mu}$ is expected
to be the same as the exponent for
ingoing matrix elements of matrix $G$.
Indeed, for the ingoing matrix elements of
Cambridge and Oxford networks 
we find respectively the exponents
$\mu = 2.12 \pm 0.03$ and $2.06 \pm 0.02$ 
(see curves in Fig.~\ref{fig3}).
For ingoing links distribution of
Cambridge and Oxford networks we obtain respectively
$\tilde{\mu} =2.29 \pm 0.02$ and  $\tilde{\mu} =2.27 \pm 0.02$
which are close to the usual WWW value 
$\tilde{\mu} \approx 2.1$.
Thus we can say that
for the WWW type networks we have $\mu \approx \tilde{\mu}$.
In contrast  the exponent
$\mu$ for DNA Google matrix elements
gets significantly larger value
$\mu \approx 5$. This feature marks a significant difference
between DNA and WWW networks.

For DNA we see that 
there is a certain curvature
in addition to a linear decay in log-log scale.
From one side, all species are close to a unique
universal decay curve which describes the 
distribution of ingoing matrix elements $g_{s}$
(there is a more pronounced deviation for DR
which does not belong to  mammalian species).
However, from other side we see visible differences
between distributions of various species
(e.g. non mammalian DR case has the largest deviation from others
 mammalian species).
We will discuss the links between $\mu$
and the exponent $\beta$ of
PageRank algebraic decay $P(K) \propto 1/K^{\beta}$
in next sections.

\subsubsection*{Spectrum of DNA Google matrix}
\label{ssec2}\ 

\begin{figure*}[!ht] 
\begin{center} 
\includegraphics[width=0.7\columnwidth]{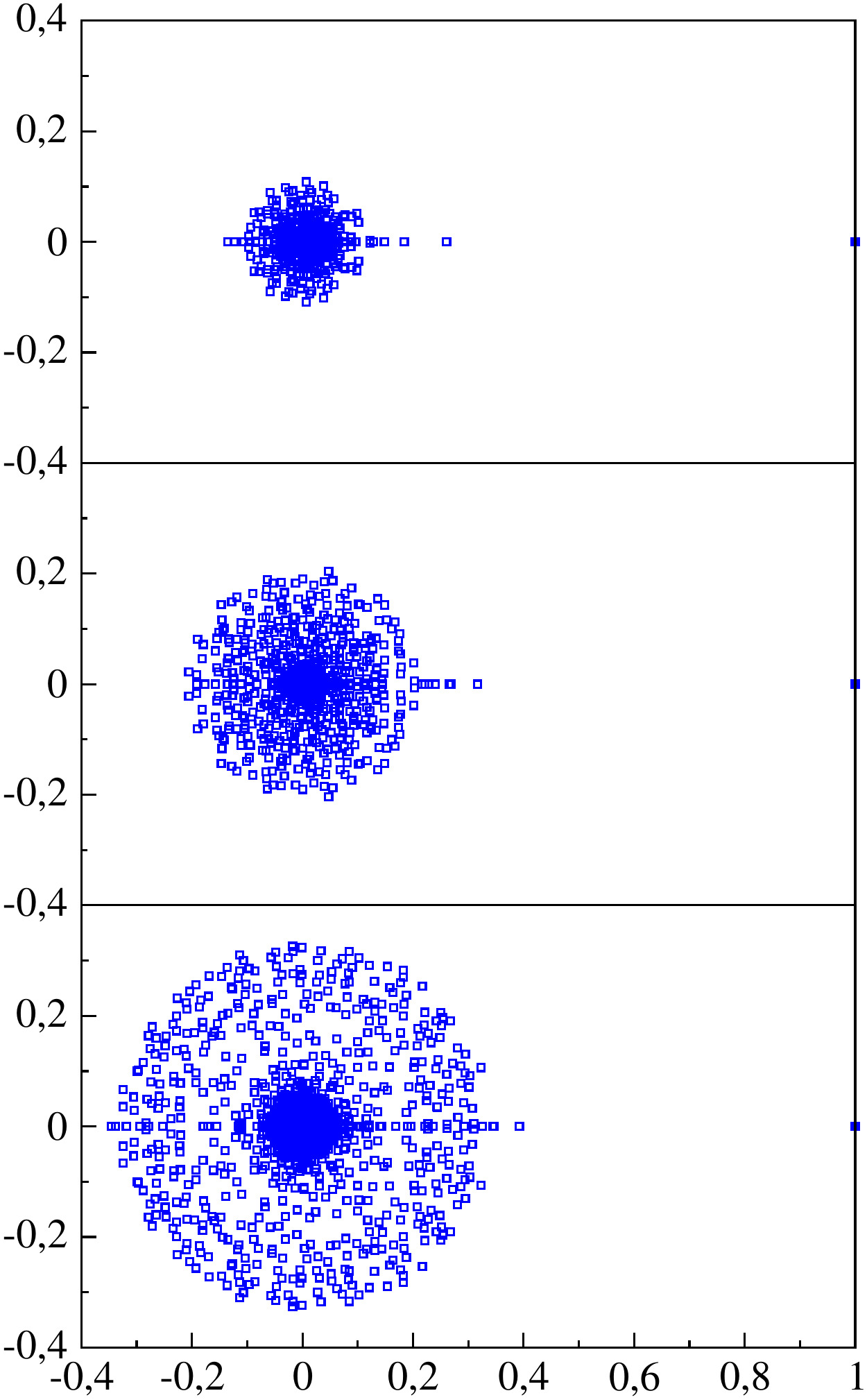}
\caption {\baselineskip 14pt
Spectrum of eigenvalues in the complex plane 
$\lambda$ for DNA Google matrix of Homo sapiens (HS)
shown for words of $5, 6, 7$ letters (from top to bottom).}
\label{fig4}\label{figure4} 
\end{center}
\end{figure*}

\begin{figure*}[!ht] 
\begin{center} 
\includegraphics[width=0.5\columnwidth]{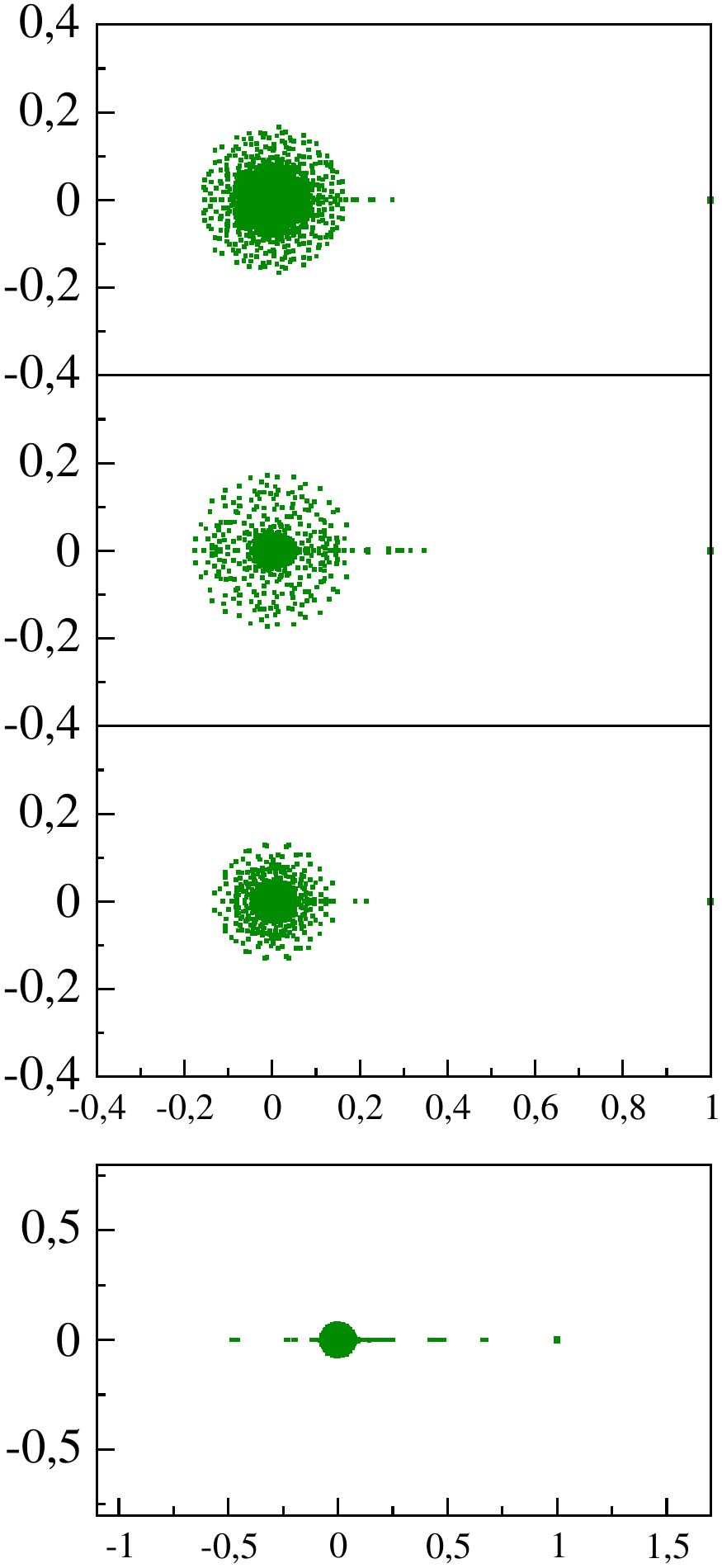}
\caption {\baselineskip 14pt
Spectrum of eigenvalues in the complex plane  $\lambda$ 
for DNA Google matrix of
of bull BT, dog CF, elephant LA, zebrafish DR
shown for words of $6$ letters  (from top to bottom).}
\label{fig5}\label{figure5} 
\end{center}
\end{figure*}

The spectrum of eigenstates of DNA Google matrix $G$ of $HS$
is shown in Fig.~\ref{fig4} for words of $m=5,6,7$ letters
and matrix sizes $N=4^m$. The spectra for DNA sequences of 
bull BT, dog CF, elephant LA and zebrafish DR are shown in Fig.~\ref{fig5}
for words of $m=6$ letters. The spectra and eigenstates are obtained 
by direct numerical diagonalization of matrix $G$
using LAPACK standard code.

In all cases the spectrum has a large gap which separates
eigenvalue $\lambda=1$ and all other eigenvalues 
with $|\lambda|<0.5$ (only for non mammalian DR case we have
a small group of eigenvalues within
$0.5 < |\lambda| < 0.75$). This is drastically different from the spectrum
of WWW and other type networks
which usually have no gap in the vicinity of $\lambda=1$
(see e.g. \cite{ukuniv,twitter},\cite{wikispectr}).
In a certain sense the DNA $G$ spectrum is similar to the spectrum
of randomized WWW networks and the spectrum of $G$ 
of the Albert-Bara\'asi network model discussed in \cite{ggs},
but the properties of the PageRank vector are rather 
different as we will see below.

Visually the spectrum is mostly similar between HS and CF
having approximately the same radius of circular cloud
$|\lambda| < \lambda_c \approx 0.2$. For DR this radius
is the smallest with $\lambda_c \approx 0.1$.
Thus the spectrum of $G$ indicates the difference
between mammalian and non mammalian sequences.
For HS the increase of the word length 
$m=5; 6; 7$ leads to an increase of 
$\lambda_c \approx 0.1; 0.2; 0.35$.
For $m=7$ the number of  nonzero matrix elements $G_{ij}$ is
close to $N^2$ and thus on average we have only 
about  $L/(m N^2) \approx 8$ transitions 
per each element. This determines 
an approximate limit
of reliable statistical computation of
matrix elements $G_{ij}$ for available
HS sequence length $L$. For HS at $m=6$
we verified that two halves of the whole sequence $L$
still give practically the same 
spectrum with a relative accuracy
of $\Delta \lambda /\lambda \approx 0.01 $
for eigenvalues in the main part
of the cloud at $\lambda_c/3 <|\lambda| < \lambda_c$.
This means that the spectrum presented in Figs~\ref{fig4},\ref{fig5}
is statistically stable at the values of $L$ used in this work.

We also constructed the Google matrix $G^*$  by inverting 
the direction of  transitions
$T_{ij} \rightarrow T_{ji}$ and then normalizing
sum of all elements in each column to unity.
This procedure is also equivalent to moving
along the sequence, from word to word,
not from left to right but from right to left.
We note that for WWW and other networks
such a  matrix with inverted direction of links
was used to obtain the CheiRank vector
(which is the PageRank vector of matrix $G^*$).
Due to the inversion of links the CheiRank vector
highlights very communicative nodes
 \cite{ukuniv,2dmotor},\cite{twitter,wikispectr}.
In our case the spectrum of $G$ and $G^*$ are identical.
As a result the probability distributions  of
PageRank and CheiRank vectors are the same.
This is due to some kind of detailed balance principle:
we count only transitions between nearby words in 
a DNA sequence and the direction of displacement along the
sequence does not affect the average transition probabilities
so that $T_{ij}=T_{ji}$ (up to statistical fluctuations).
In a certain sense this situation is similar to the
case of Ulam networks in symplectic maps
where the conservation of phase space area
leads to the same properties of $G$ and $G^*$
\cite{ulamfrahm,acat}.

We tried to test if a random matrix model can reproduce the
distribution of eigenvalues in $\lambda$ plane.
With this aim we generated random matrix elements $G_{ij}$
with exactly the same distribution $N_g$ as for HS case at $m=6$
(see  Fig.~\ref{fig2}). However, in this random model
we found all eigenvalues homogeneously distributed in the
radius $\lambda_c \approx 0.07$ being significantly smaller
compared to the real data. Also in this case the PageRank
probability $P(K)$ changes only by 30\% in the whole
range $1 \leq K \leq N$ being absolutely different from the
real data (see next section). Thus the construction
of random matrix models which are able to produce
results similar to the real data remains as a task for future 
investigations.

\subsubsection*{PageRank properties of various species}
\label{ssec3}\ 

\begin{figure*}[!ht] 
\begin{center} 
\includegraphics[width=0.9\columnwidth]{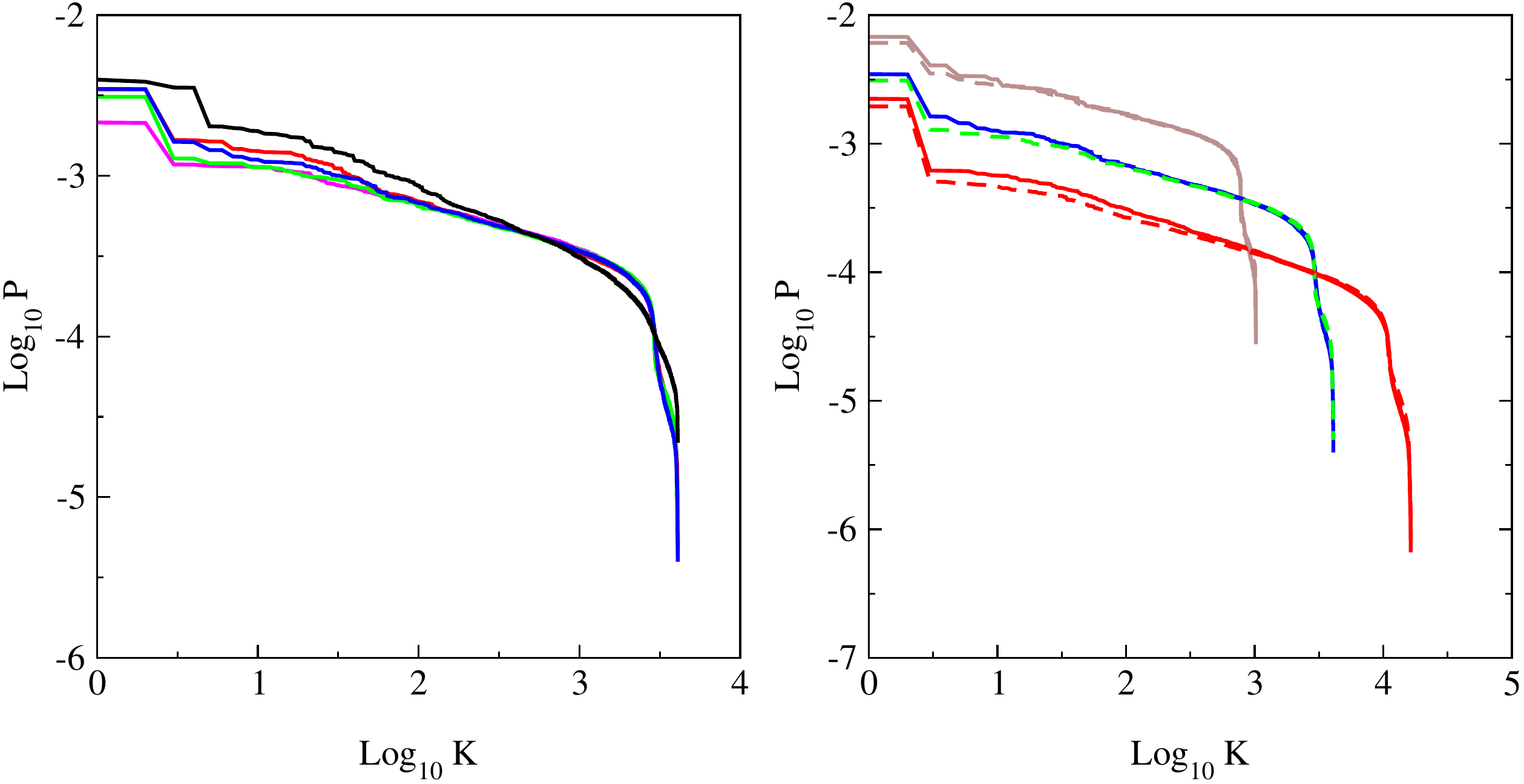}
\caption {\baselineskip 14pt
Dependence of PageRank probability $P(K)$ on PageRank index $K$. 
\emph{Left panel :} 
Data for different species for word length of  6-letters: bull BT (magenta), 
dog CF (red), elephant LA (green), 
Homo sapiens HS (blue) and zebrafish DR (black). 
\emph{Right panel :} Data for HS (full curve) and LA (dashed curve)
for word length $m=5$  (brown), $6$ (blue/green), $7$ (red).} 
\label{fig6}\label{figure6} 
\end{center}
\end{figure*}

By numerical diagonalization of the Google matrix we determine
the PageRank vector $P(K)$ at $\lambda=1$ and several other
eigenvectors with maximal values of $|\lambda|$.
The dependence of probability $P$ on index $K$
is shown in Fig.~\ref{fig6} for various species and
different word length $m$. The probability $P(K)$
describes  the steady state of random walks on the Markov chain
and thus it gives the frequency of appearance of various
words of length $m$ in the whole sequence $L$.
The frequencies or probabilities 
of words appearance in the sequences
have been obtained in \cite{buldyrev1995}
by a direct counting of words along the sequence 
(the available sequences $L$ were shorted at that times).
Both methods are mathematically equivalent 
and indeed our distributions $P(K)$ are in a good agreements
with those found in \cite{buldyrev1995}
even if now we have a significantly better statistics.

The decay of $P$ with $K$
can be approximately described by a power law
$P \sim 1/K^{\beta}$. Thus for example for HS sequence at $m=7$
we find  $\beta=0.357 \pm 0.003$
for the fit range $1.5 \leq \log_{10} K \leq 3.7$
that is rather close to the exponent found in \cite{buldyrev1995}.
Since on average the PageRank probability is proportional
to the number of ingoing links, or the sum
of ingoing matrix elements of $G$, one has the relation between
the exponent of PageRank $\beta$ and exponent of ingoing
links (or matrix elements): $\beta = 1/(\mu -1)$
\cite{meyerbook,ukuniv},\cite{donato,upfal},\cite{zzs}.
Indeed, for the HS DNA case at $m=7$ we have
$\mu=4.48$ that gives $\beta = 0.29$ being close to the
above value of $\beta =0.357$ obtained from the direct fit
of $P(K)$ dependence. We think that the agreement is
not so perfect since there is a visible curvature in the
log-log plot of $N_s$ vs $g_{s}$ in Fig.~\ref{fig3}.
Also due to a small value of $\beta$ the variation range
of $P$ is not so large that reduces the accuracy 
of the numerical fit even if a formal statistical error
is relatively small compared to a visible systematic
nonlinear variations. In spite of this only approximate agreement
we should say that in global the relation between $\beta$ and $\mu$
works correctly. In average we find for DNA network 
the value of $\mu \approx 5$ being 
significantly larger than for the WWW networks with  $\tilde{\mu} \approx 2.1$
\cite{meyerbook}. 
This   gives a significantly smaller value $\beta \approx 0.25$ for DNA case
comparing to the usual WWW value $\beta \approx 0.9$
(we note that the randomized WWW networks and the Albert-Barab\'asi model
have $\beta \approx 1$ \cite{ggs}).
The relation between $\beta$ and $\mu$ also works
for the DR DNA case at $m=6$ with $\mu =4.04$
that gives $\beta = 0.33$ being in a satisfactory agreement
with the fit value $\beta = 0.426 $
found from $P(K)$ dependence of Fig.~\ref{fig6}.

At $m=6$ we find for our species the following values
of  exponent $\beta = 0.273 \pm 0.005$ (BT),
$0.340 \pm 0.005$ (CF), $0.281 \pm 0.005$ (LA), $0.308 \pm 0.005$ (HS),
$0.426 \pm 0.008$ (DR) in the range $1 \le \log_{10}K \leq 3.3$.
There is a relatively small variation of $\beta$
between various mammalian species.
The data of Fig.~\ref{fig6} for HS show that 
the value of $\beta$ remains stable with the increase 
of  word length. These observations are similar to those
made in \cite{buldyrev1995}.

\subsubsection*{PageRank proximity between species}
\label{ssec4}\ 

\begin{figure*}[!ht] 
\begin{center} 
\includegraphics[width=0.9\columnwidth]{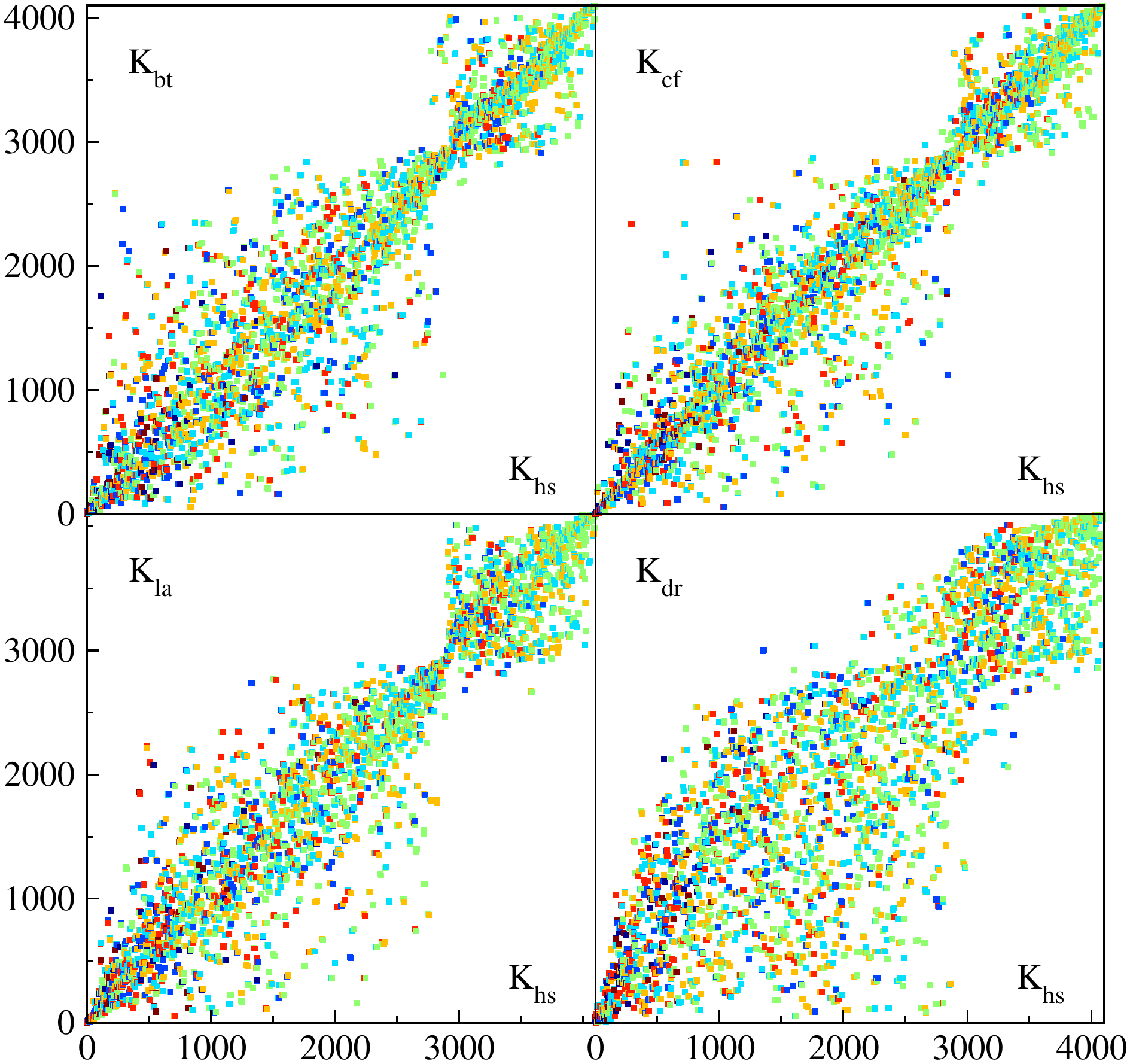}
\caption {\baselineskip 14pt
PageRank proximity $K-K$ plane diagrams 
for different species in comparison with Homo sapiens: 
$x$-axis shows PageRank index $K_{hs}(i)$ of a word $i$ and
$y$-axis shows PageRank index of the same word $i$ 
with $K_{bt}(i)$ of bull, $K_{cf}(i)$ of dog, 
$K_{la}(i)$ of elephant and $K_{dr}(i)$ of zebrafish;
here the word length is $m=6$. 
The colors of symbols marks the purine content in a word $i$ 
(fractions of letters $A$ or $G$ in any order);
the color varies from red at maximal content, via brown, 
yellow, green, light blue,
to blue at minimal zero content.}
\label{fig7}\label{figure7} 
\end{center}
\end{figure*}

\begin{figure*}[!ht] 
\begin{center} 
\includegraphics[width=0.9\columnwidth]{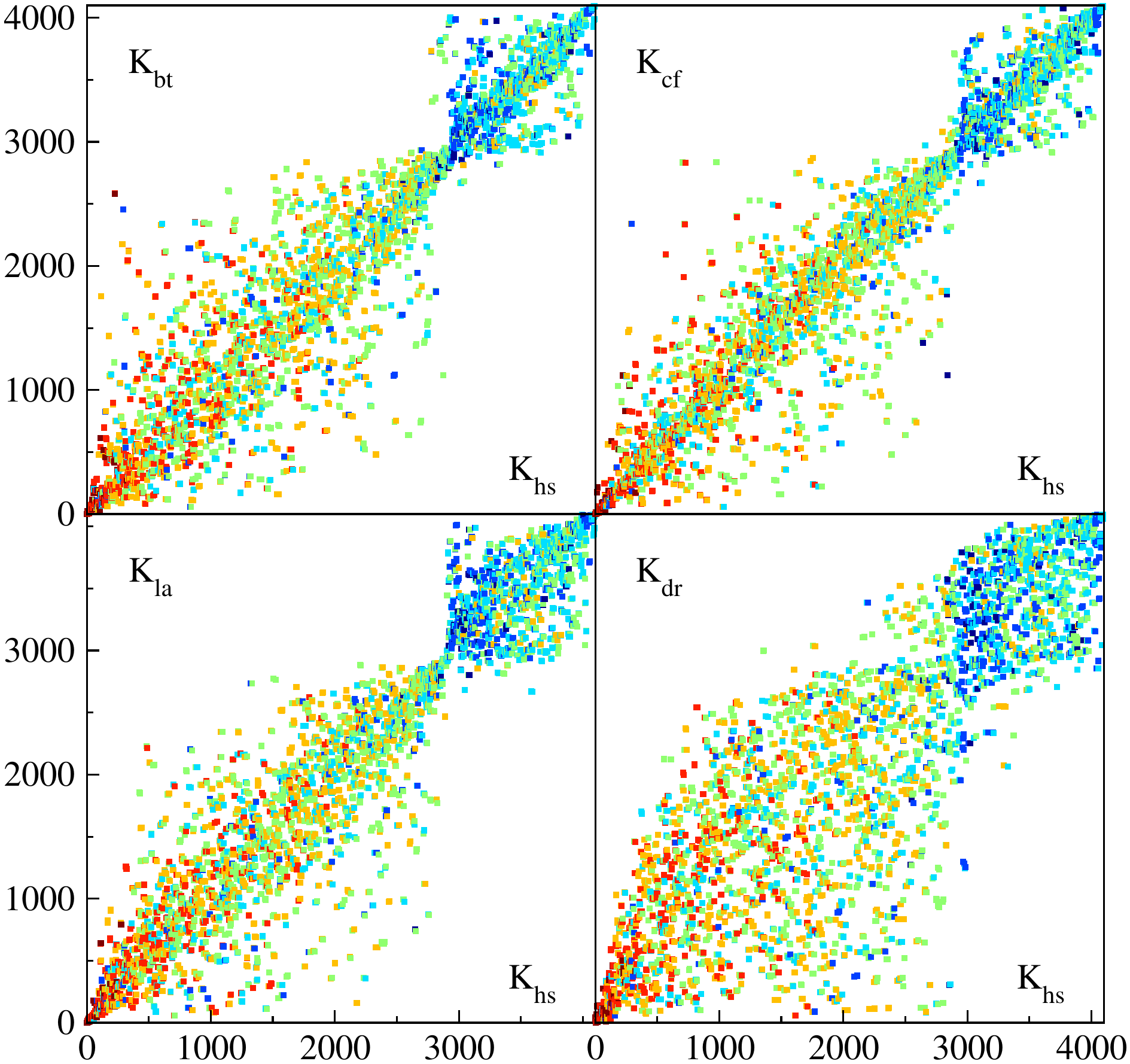}
\caption {\baselineskip 14pt
Same as in Fig.~\ref{fig7} but now the color
marks the fraction of of letters $A$ or $T$ in any order 
in a word $i$ with red at maximal content
and blue at zero content.}
\label{fig8}\label{figure8} 
\end{center}
\end{figure*}

\begin{figure*}[!ht] 
\begin{center} 
\includegraphics[width=0.9\columnwidth]{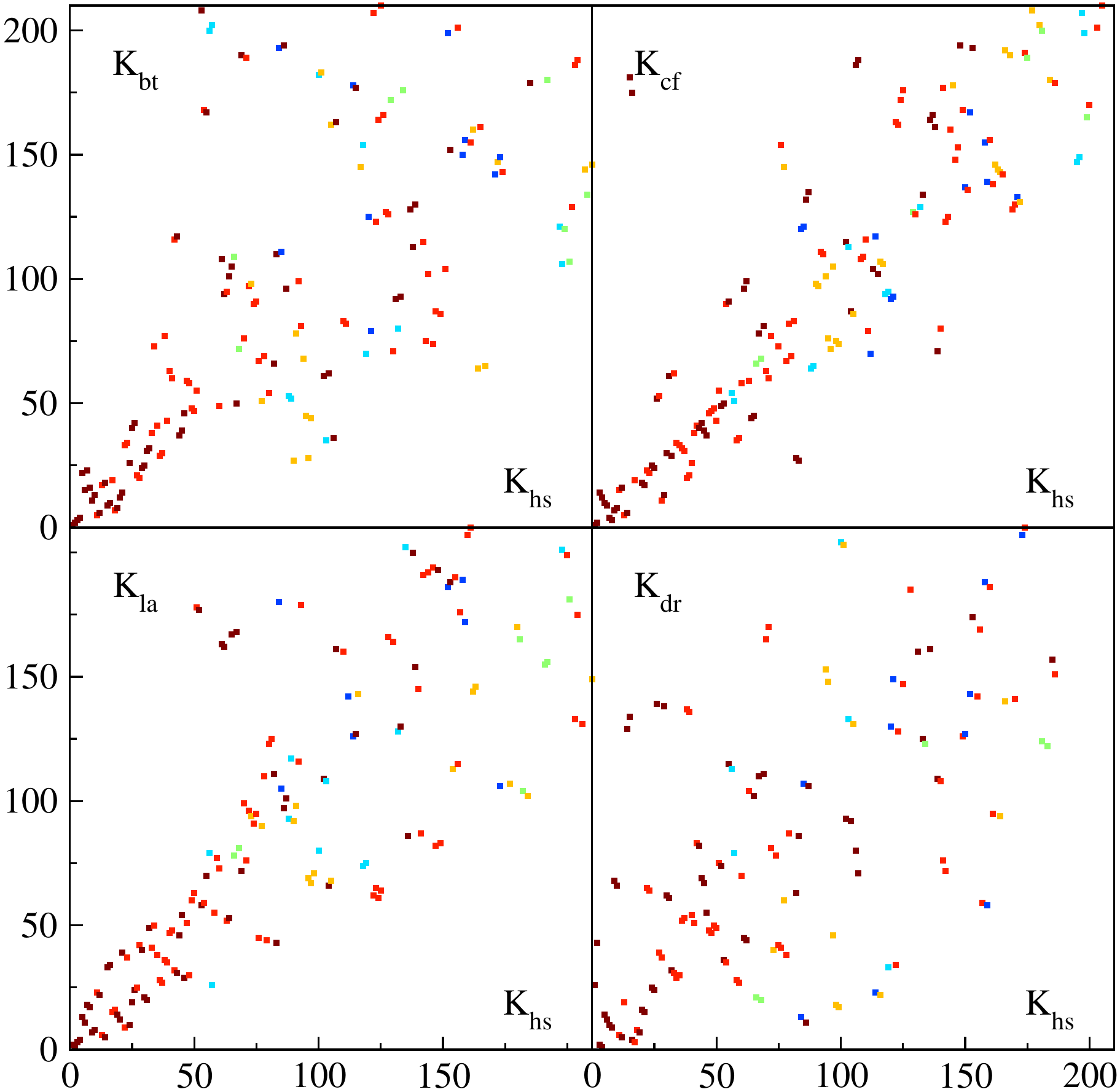}
\caption {\baselineskip 14pt
Zoom of the PageRank proximity $K-K$ diagram of Fig.~\ref{fig8}
for the range $1 \leq K \leq 200$ with the same color for $A$ or $T$
content.}
\label{fig9}\label{figure9} 
\end{center}
\end{figure*}

The top ten 6-letters words,
with largest probabilities $P(K)$, are given for all studied species 
in Table~\ref{table1}. Two top words are identical for BT, CF, HS.
To see a similarity between species on a global scale
it is convenient to plot
the PageRank index $K_{s}(i)$ of a given species $s$
versus the index $K_{hs}(i)$ of HS for the same word $i$.
For identical sequences one should have all points
on diagonal, while the deviations from diagonal
characterize the differences between species.
The examples of such PageRank proximity $K-K$ diagrams
are shown in Figs.~\ref{fig7},\ref{fig8}
for words at $m=6$. 
A zoom of data on a small scale at the range $1\leq K \leq 200$
is shown in Fig.~\ref{fig9}. 
A visual impression is that
CF case has less deviations from HS rank
compared to BT and LA. The non-mammalian DR case
has most strong deviations from HS rank.
For BT, CF and LA cases we have a significant
reduction of deviations from diagonal around
$K \approx 3N/4$. This effect is also visible
for DR case even if being less pronounced.
We do not have explanation for this observation.

The fraction of purine letters $A$ or $G$
in a word of $m=6$ letters 
is shown by color in Fig.~\ref{fig7} for all words
ranked by PageRank index $K$.
We see that these letters are approximately 
homogeneously distributed over the whole range
of $K$ values.
In contrast to that the distribution of letters $A$ or $T$
is inhomogeneous in $K$: their fraction is 
dominant for $1\leq K < N/4$,
approximately homogeneous for $N/4 \leq K \leq 3N/4$
and is close to zero for $ 3N/4 < K \leq N$ (see Fig.~\ref{fig8}).
We find that in the whole HS sequence the fractions $F_{a,c,g,t}$ of
$A,C,G,T$ are respectively $0.276596, 0.192576, 0.192624, 0.276892$
(and $F_n=0.061312$ for undetermined $N_l$).
Thus we have the fraction of $A,G$ being close to 
$1/2 \approx (F_a+F_g)/(1-F_n) = 0.499867$ and the fraction of $A,T$ being
$(F_a+F_t)/(1-F_n) = 0.589640 > 0.5$. Thus it is more probable to
have $A$ or $T$ in the whole sequence
that can be a possible origin of the inhomogeneous distribution
of $A$ or $T$ along $K$ and large fraction of
$A$, $T$ at top PageRank positions.

\begin{figure*}[!ht] 
\begin{center} 
\includegraphics[width=0.9\columnwidth]{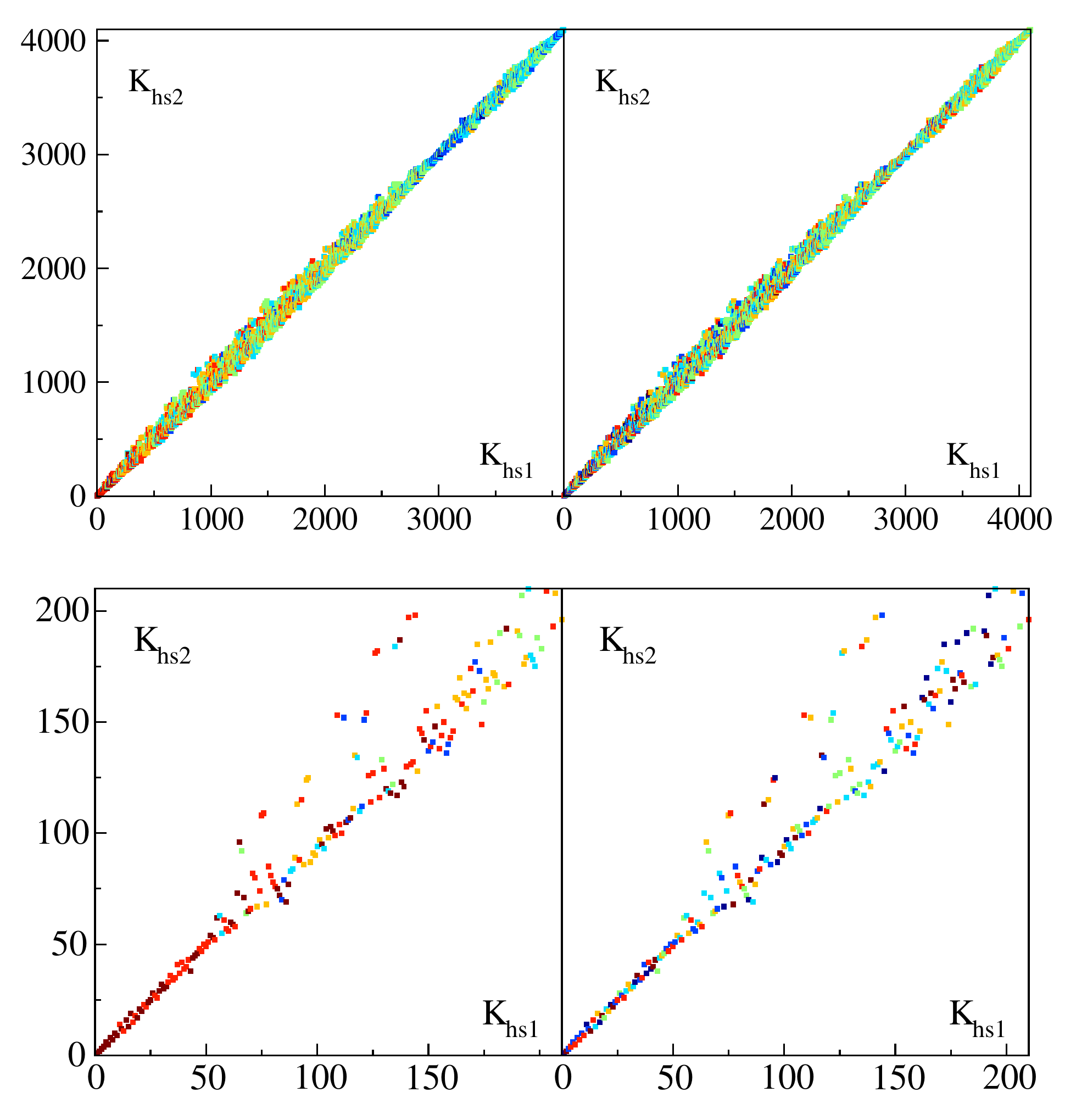}
\caption {\baselineskip 14pt
PageRank proximity $K-K$ diagram 
of Homo sapiens $HS2$ versus Homo sapiens $HS1$
at $m=6$ (see text for details).
Top panels show the content of $A,T$ (left) and 
$A,G$ (right) in the same way as in Fig.~\ref{fig8}
and Fig.~\ref{fig7} respectively.
Bottom panels show zoom of top panels.}
\label{fig10}\label{figure10} 
\end{center}
\end{figure*}

The whole HS sequence used here is composed from 
5 humans with individual length $L_i \approx 3 \cdot 10^9 \approx L/5$.
We consider the first and last fifth parts of the whole
sequence $L$ separately thus forming two independent 
sequences HS1 and HS2 of two individuals.
We determine for the the corresponding 
PageRank indexes $K_{hs1}$ and $K_{hs2}$
and show their PageRank proximity diagram in Fig.~\ref{fig10}.
In this case the points are much closer to diagonal
compared to the case of comparison of HS with other species.

To characterize the proximity between different species
or different HS individuals we compute 
the average dispersion 
$\sigma(s_1,s_2) = \sqrt{\sum_{i=1}^N (K_{s_1}(i)- K_{s_2}(i))^2)/N}$
between two species (individuals) $s_1$ and $s_2$.
Comparing the words with length $m=5,6,7$ we find
that the scaling  $\sigma \propto N$
works with a good accuracy (about 10\% when $N$
is increased by a factor 16). To represent the result
in a form independent of $m$ we compare the values of
$\sigma$ with the corresponding random model
value $\sigma_{rnd}$. This value is computed assuming
a random distribution of $N$ points in 
a square $N \times N$ when  only one point appears 
in each column and each line
(e.g. at $m=6$ we have $\sigma_{rnd} \approx 1673$ 
and $\sigma_{rnd} \propto N$). The dimensionless dispersion
is then given by $\zeta(s_1,s_2)=\sigma(s_1,s_2)/\sigma_{rnd}$.
From the ranking of different species we obtain the following values
at $m=6$:
$\zeta(CF,BT)=0.308$; $\zeta (LA,BT)=0.324$,  $\zeta(LA,CF)=0.303$;
$\zeta(HS,BT)=0.246$, $\zeta(HS,CF)=0.206$, $\zeta(HS,LA)=0.238$;
$\zeta(DR,BT)=0.425$, $\zeta(DR,CF)=0.414$, $\zeta(DR,LA)=0.422$,
$\zeta(DR,HS)=0.375$ (other $m$ have similar values).
According to this statistical analysis of PageRank proximity
between species we find that $\zeta$ value is minimal between
CF and HS showing that these are two most similar 
species among those considered here.

For two HS individuals we find
$\zeta(HS1,HS2) = 0.031$ being significantly smaller
then the proximity correlator 
between different species.
We think that this  PageRank proximity correlator
$\zeta$ can be useful as a quantitative measure
of statistical proximity between various species.

Finally, in Table~\ref{table2} we give for all species the words
of 6 letters with the 10 minimal PageRank probabilities.
Thus for HS the less probable is the word TACGCG 
corresponding to  two amino acids Tyr and Ala.
In general the ten last words are mainly composed of C and G
even if the letters A and T still have small but nonzero 
weight. The last two words are the same for mammalian species
but they are different for DR sequence.

\subsubsection*{Other eigenvectors of G}
\label{ssec5}\

\begin{figure*}[!ht] 
\begin{center} 
\includegraphics[width=0.9\columnwidth]{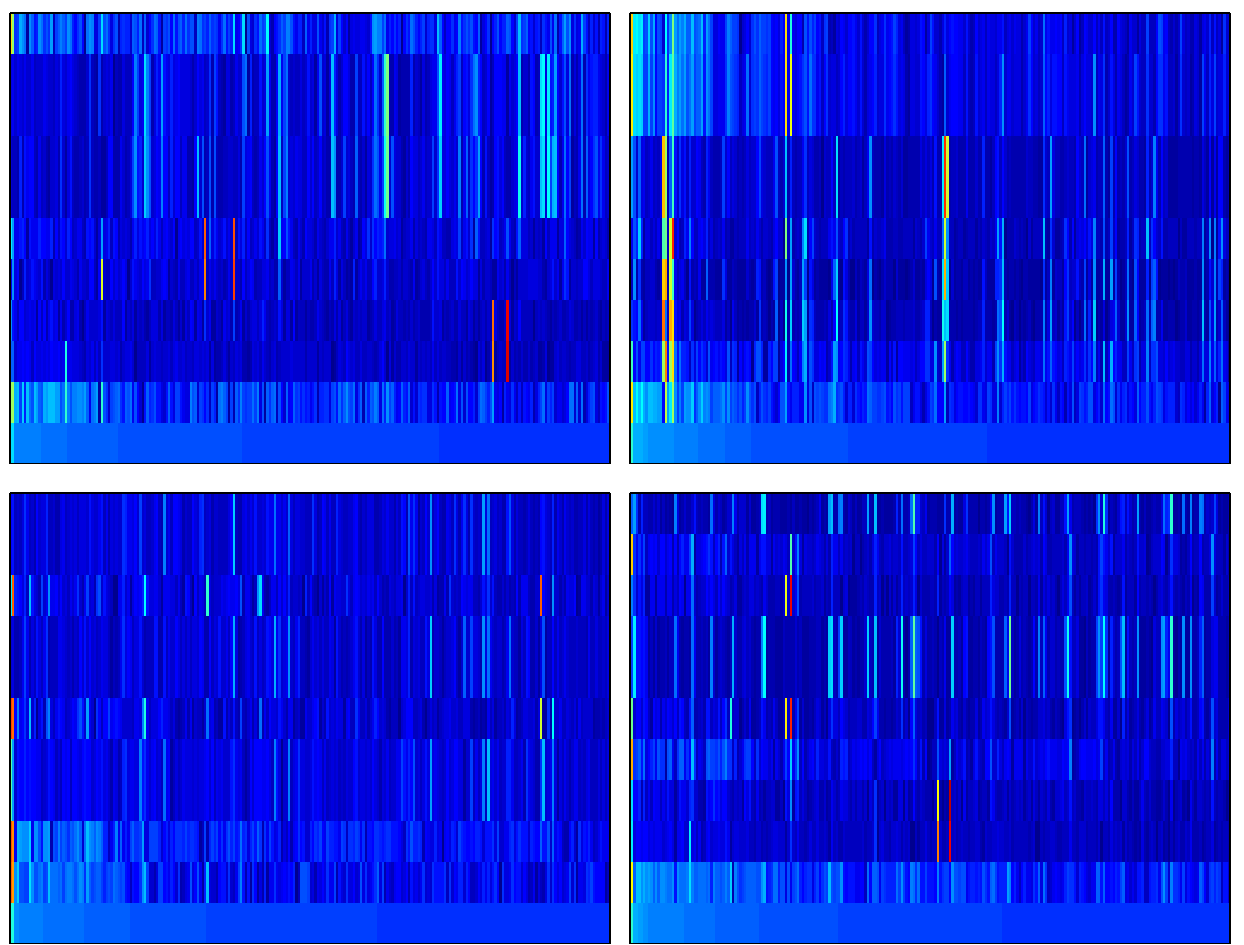}
\caption {\baselineskip 14pt
Dependence of eigenstates amplitude
$|\psi_i(K)|$ on PageRank index $K$ in $x$-axis
and eigenvalue index $i$ in $y$-axis  for largest ten
eigenvalues $|\lambda_i|$ counted by
$i$ from $i=1$ at $|\lambda_1|=1$
to $i=10$ at $ |\lambda_{10}| \approx 0.2$.
The range $1 \leq K \leq 250$ is shown 
with PageRank vector for a given species
at the bottom line of each panel.
For each species in each panel the color is proportional
to $\sqrt{|\psi_i(j)|}$ changing from blue at zero 
to red at maximal amplitude value which is
close to unity in each panel.
The panels show the species: bull BT (top left),
 dog CF (top right), elephant LA (bottom left),
Homo sapiens HS (bottom right).}
\label{fig11}\label{figure11} 
\end{center}
\end{figure*}

The properties of 10 eigenstates $\psi_i(j)$ of DNA Google matrix
with largest modulus of eigenvalues $|\lambda_i|$
 are analyzed in Table~\ref{table3} and Fig.~\ref{fig11}.
The words $W_i$ at the maximal amplitude $|\psi_i(j)|$
are presented for all species in Table~\ref{table3}.
We see that in general these words $W_i$ are rather
different from the top PageRank word $W_1$
(some words appear in pairs since there are pairs of
complex conjugated values $\lambda_i=\lambda_i^*$).

The probability of the above top 10 eigenstates
as a function of PageRank index $K$ are shown in Fig.~\ref{fig11}.
We see that the majority of the vectors,
different from the PageRank vector,
have well localized peaks
at relatively large values $K > 50$.
This shows that in the DNA network there are some modes
located on certain specific patterns
of words. 

To illustrated the localized structure 
of eigenmodes $\psi_i(j)$ for HS case
at $m=6$ we compute the inverse participation ratio
$\xi_{i}=(\sum_j |\psi_i(j)|^2)^2/\sum_j |\psi_i(j)|^4$
which gives an approximate number of nodes on which the main probability of
an eigenstate $\psi_i(j)$ is located
(see e.g. \cite{ukuniv,twitter,ggs}).
The obtained values are
$\xi_i=385.26$, $16.37$, $2.07$, $1.72$, $2.23$, $3.19$,
$77.43$, $77.43$, $2.33$, $2.06$ 
for $i=1,...10$ respectively.
We see that for $i>1$ we have significantly smaller $\xi$
values compared to the case of PageRank vector with 
a large $\xi_1$. This supports the conclusion about 
localized structure of a large fraction of eigenvectors of $G$.

In \cite{wikispectr} on an example of Wikipedia
network it is shown that the eigenstates with 
relatively large
$|\lambda|$ select specific communities of the
network. The detection of communities in complex networks
is now an active research direction \cite{fortunato}.
We expect that the eigenmodes of G matrix
can select specific words of 
bioniformatic interest. However,
a detailed analysis of words from 
eigenmodes remains for further more detailed investigations.

\section*{Discussion}

In this work we used long DNA sequences of various species 
to construct from them the Markov process describing
the probabilistic transitions between words of up to 7 letters length. 
We construct the Google matrix of such transitions 
with the size up to $4^7$ and analyze
the statistical properties of its matrix elements.
We show that for all 5 species, studied in this work,
the matrix elements of significant amplitude
have a power law distribution with the exponent
$\nu \approx 2.5$ being close to the exponent
of outgoing links distribution typical for WWW 
and other complex directed networks with $\tilde{\nu} \approx 2.7$.
The distribution of significant values of the sum of ingoing 
matrix elements of $G$ 
is also described by a power law with the exponent
$\mu \approx 5$ which is significantly larger than 
the corresponding exponent for WWW networks 
with $\tilde{\mu} \approx 2.1$. We show 
that similar to the WWW networks the
exponent $\mu$ determines the exponent 
$\beta=1/(\mu-1) \approx 0.25$ of the
algebraic PageRank decay 
which is significantly smaller then its value for WWW 
networks with $\beta \approx 0.9$. 
The PageRank decay is similar to the frequency decay
of various words studied previously in \cite{buldyrev1995}.
It is interesting to note
that the value $\mu-1$ is close to the exponent 
of Poincar\'e recurrences decay which has a value 
close to 4 \cite{dnafrahm} 
(even if we cannot derive a direct mathematical
relation between them).

Using PageRank vectors of various species 
we introduce the PageRank proximity correlator
$\zeta$ which allows to measure in a quantitative way
the proximity between different species.
This parameter remains stable in respect to variation of
the word length. 

The spectrum of the Google matrix is determined
and it is shown that it is characterized by a significant
gap between $\lambda=1$ and other eigenvalues.
Thus, this spectrum is qualitatively different from the WWW
case where the gap is absent at the damping factor $\alpha=1$. 
We show that the eigenmodes with largest values of
$|\lambda|<1$ are well localized on specific words
and we argue that the words corresponding to 
such localized modes can play an interesting role in
bioinformatic properties of DNA sequences.

Finally we would like to trace parallels between
the Google matrix analysis of words in DNA sequences and the 
small world properties of human language. 
Indeed, it is known that the frequency of words 
in natural languages follows a power law Zipf
distribution with the exponent $\beta \approx 1$ \cite{zipf}. 
The parallels between words distributions
in DNA sequences and statistical linguistics 
were already pointed in \cite{buldyrev1995}.
The analysis of degree distributions of undirected networks of 
words in natural languages was found to follow 
a power law with an exponent $\nu_l \approx 1.5 - 2.7$ \cite{sole}
being not so far from the one found here
for the matrix elements distribution.
It is argued that the language evolution 
plays an important role in the formation
of such a distribution in languages \cite{dorogovtsev}.
The parallels between linguistics and DNA sequence complexity are
actively discussed in bioinformatics \cite{trifonov1,trifonov2}.
We think that the Google matrix analysis 
can provide new insights in the construction and 
characterization of information flows on DNA sequence networks 
extending recent  steps  done in \cite{trifonov3}.

In summary, our results show that the distributions
of significant matrix elements are similar to those
of the scale-free type networks
like WWW, Wikipedia and linguistic networks.
In analogy with lingusitic networks
it can be useful to go from words network
analysis to a more advanced functional level of 
links inside sentences that may be
viewed as a network of links between amino acids
or more complex biological constructions.

\section*{Acknowledgments}

We thank K.M.Frahm for useful discussions and help in collection
of DNA sequences from \cite{dnafrahm} which are studied here.
This research is supported in part by the EC FET Open project 
``New tools and algorithms for directed network analysis''
(NADINE $No$ 288956); 
VK is supported by CNRS - Region Midi-Pyr\'en\'ees grant.
We also acknowledge
the France-Armenia collaboration grant CNRS/SCS No. 24943
(IE-017) on “Classical and quantum chaos”.

\section*{Supporting Information}

Supplementary methods, references, tables, sequences data
and figures are available at:\\
http://www.quantware.ups-tlse.fr/QWLIB/dnagooglematrix/

\newpage$\phantom{.}$

\newpage$\phantom{.}$

\newpage$\phantom{.}$

\newpage$\phantom{.}$

\newpage$\phantom{.}$

\newpage$\phantom{.}$

\newpage$\phantom{.}$

\newpage$\phantom{.}$

\newpage$\phantom{.}$

\newpage$\phantom{.}$

\newpage$\phantom{.}$

\newpage$\phantom{.}$

\begin{table}[!ht]
\caption{Top ten PageRank entries at DNA word length $m=6$ 
for  species: bull BT, dog CF, 
elephant LA, Homo sapiens HS  and zebrafish DR.}
\begin{center}
\resizebox{7cm}{!}{
\begin{tabular}{|c|c|c|c|c|}
  \hline
  BT & CF & LA & HS & DR \\
  \hline
TTTTTT&	TTTTTT&	AAAAAA&	TTTTTT& ATATAT \\
AAAAAA&	AAAAAA&	TTTTTT&	AAAAAA&	TATATA \\
ATTTTT&	AATAAA&	ATTTTT&	ATTTTT&	AAAAAA \\
AAAAAT&	TTTATT&	AAAAAT&	AAAAAT&	TTTTTT \\
TTCTTT&	AAATAA&	AGAAAA&	TATTTT&	AATAAA \\
TTTTAA&	TTATTT&	TTTTCT&	AAAATA&	TTTATT \\
AAAGAA&	AAAAAT&	AAGAAA&	TTTTTA&	AAATAA \\
TTAAAA&	ATTTTT&	TTTCTT&	TAAAAA&	TTATTT \\
TTTTCT&	TTTTTA&	TTTTTA&	TTATTT&	CACACA \\
AGAAAA&	TAAAAA&	TAAAAA&	AAATAA&	TGTGTG \\
  \hline
\end{tabular}}
\end{center}
\label{table1}
\end{table}

\newpage$\phantom{.}$

\begin{table}[!ht]
\caption{Ten words with minimal PageRank probability
given at $m=6$ for species:
bull BT, dog CF, elephant LA, Homo sapiens HS  and zebrafish DR.
Here the top row is the last PageRank entry, 
bottom is the tenth one from the end
of PageRank.}
\begin{center}
\resizebox{7cm}{!}{
\begin{tabular}{|c|c|c|c|c|}
  \hline
  BT & CF & LA & HS & DR \\
  \hline
CGCGTA&	TACGCG&	CGCGTA&	TACGCG& CCGACG \\
TACGCG&	CGCGTA&	TACGCG&	CGCGTA&	CGTCGG \\
CGTACG&	TCGCGA&	ATCGCG&	CGTACG&	CGTCGA \\
CGATCG&	CGTACG&	TCGCGA&	TCGACG&	TCGACG \\
ATCGCG&	CGATCG&	CGCGAT&	CGTCGA&	TCGTCG \\
CGCGAT&	CGAACG&	GTCGCG&	CGATCG&	CCGTCG \\
TCGACG&	CGTTCG&	CGATCG&	CGTTCG&	CGACGG \\
CGTCGA&	TCGACG&	CGCGAC&	CGAACG&	CGACCG \\
CGTTCG&	CGTCGA&	TCGCGC&	CGACGA&	CGGTCG \\
TCGTCG&	ACGCGA&	ACGCGA&	CGCGAA&	CGACGA \\
  \hline
\end{tabular}}
\end{center}
\label{table2}
\end{table}

\newpage$\phantom{.}$

\begin{table}[!ht]
\caption{Words $W_i$ corresponding to the maximum value 
of eigenvector modulus 
$w_i=max_{j}(|\psi_i(j)|)$ for species 
 bull BT, dog CF, 
elephant LA, Homo sapiens HS  and zebrafish DR, which are 
shown in dark red in Fig.~\ref{fig11}. 
The eigenvectors at $i=1,...,10$ correspond 
to the ten largest eigenvalues $|\lambda_1|,...,|\lambda_{10}|$ of 
the DNA Google matrix 
for DNA word length $m=6$. 
The first row $i=1$ corresponds to top PageRank entries. }
\begin{center}
\resizebox{7cm}{!}{
\begin{tabular}{|c|c|c|c|c|c|}
  \hline
 i & BT & CF & LA & HS & DR \\
  \hline
1& TTTTTT&	TTTTTT&	AAAAAA&	TTTTTT&	ATATAT \\
2& TTTTTT&	AAAAAA&	AAAAAA&	TTTTTT&	TATATA \\
3& ACACAC&	CTCTCT&	AAAAAA&	ACACAC&	ATATAT \\
4& ACACAC&	AGAGAG&	AAAAAA&	ACACAC&	TAGATA \\
5& CACACA&	CTCTCT&	AAAAAA&	TTTTTT&	ATAGAT \\
6& CACACA&	TCTCTC&	AAAAAA&	CACACA&	TATCTA \\
7& CCAGGC&	AGAGAG&	TATGAG&	TGGGAG&	ATCTAT \\
8& CCAGGC&	AGAGAG&	TATGAG&	TGGGAG&	TAGATA \\
9& CCCATG&	TGTGTG&	TTTTTT&	CACACA&	ATAGAT \\
10& CCCATG&	TGTGTG&	AGAGTA&	TTTTTT&	TATCTA \\
  \hline
\end{tabular}}
\end{center}
\label{table3}
\end{table}

\end{document}